\documentclass[pra,twocolumn,showpacs,showkeys,preprintnumbers,amsmath,amssymb]{revtex4}
\usepackage{tipa}
\usepackage{amssymb}
\usepackage{txfonts}
\usepackage{CJK}

\usepackage{hyperref}
\usepackage{graphicx}
\usepackage{epsfig}

\begin{document}\begin{CJK*}{GBK}{song}
\title{Quantum State Transfer between Distant Cavity-Optomechanics Connected via Fiber}

\author{Dai Jia-Feng, Guo Yan-Qing$^{*}$,Pei Pei\footnotetext{ $^{*}$ Corresponding author. Email:
yqguo@dlmu.edu.cn}, Zhang Xing-Yuan, Wang Dian-Fu, Huang He-Fei}
\affiliation{Department of Physics, Dalian Maritime University,
Dalian, 116026} 

\begin{abstract}
We propose a scheme for high quantum state transfer efficiency
between two distant mechanical oscillators. Through coupling
separately to two optical cavities connected by an optical fiber,
two distant mechanical oscillators achieve a transfer efficiency
closing to unity. In the process of analysis, we employed
time-varying optomechanical coupling strength as a Gaussian pulse.
\end{abstract}
\maketitle

In recent years, the cavity-optomechanical system has drawn a lot
of interest both in theoretical and experimental research. On the
one hand, it is because the optomechanical system has high optical
detection sensitivity, on the other hand, quantum optomechanical
system can use light to detect the mechanical movement in quantum
system, and generate nonclassical state light and mechanical
movement. A general resonant cavity is formed by an optical cavity
coupling with a nano mechanical oscillator. The principle is
letting the radiation pressure of the light field intra cavity
actting on the mechanical oscillator, so that the mechanical
oscillator can vibrate freely near the eigenfrequency, and the
vibration will, in turn, modulate the frequency of the light
field. This optomechanical system has ultra high sensitivity,
ultra light effective mass, ultra high quality factor, and ultra
high frequency. It is also based on these characteristics, the
optomechanical system has very high application value in precision
detection and the measurement of force and displacement
$^{[1-3]}$. With the progress of nanotechnology and the
semiconductor industry advances in materials and processes, the
size of mechanical vibrators can reach micrometer or even
nanometer. With the constantly tending to miniaturization and low
dissipation, the model of optomechanical system will provide a
feasible scheme for the detection of micromechanical vibrators,
both in technical and basic science. It will no longer be used
solely for precise measurement, but will also become a new
technology for controlling cooling $^{[4-8]}$ and amplifying
$^{[9,10]}$. In addition, the mixing optomechanical system
coupling with atoms, electrons or quantum dots is favourable to
control the optomechanical system, check the basic quantum laws
and study the quantum effects of mesoscopic or macroscopic objects
$^{[11-13]}$, such as the preparation of mechanical oscillators'
squeezed state, entanglement state and superposition state
$^{[14]}$. At the same time, the realization of controlling
quantum optomechanical system has great application value in the
processing of quantum information, and it will greatly promote the
development of quantum information $^{[15,16]}$.

In this paper, we consider the transfer of a quantum state between
two distant mechanical oscillators which are coupled to two
optical cavities separately.High efficient transmission of quantum
state is achieved by using experimental parameters $^{[16,17]}$
and controlling Gaussian coupling strength between the mechanical
oscillators and cavity.

The theoretical model used in this paper is consisted by two
optical cavities and two mechanical oscillators. Each mechanical
oscillator is coupled to an optical cavity via radiation pressure.
One cavity is connected with another optical cavity which is
coupled to the second mechanical oscillator via optical fiber(Fig.
1). At the initial
 moment, we consider a quantum state encoded onto the first mechanical oscillator and transfer
 it to the second mechanical oscillator through two middle optical cavities that are connected
 by fiber. Under the current system, we can realize the energy transfer from mechanical energy
 to mechanical energy in quantum way.

\begin{figure}
\epsfig{file=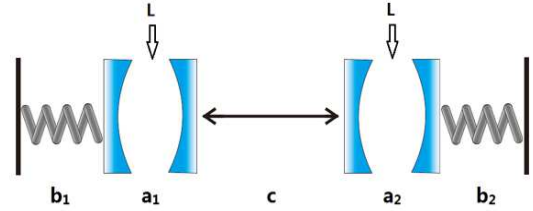, width=7cm,
height=2.8cm,bbllx=4,bblly=10,bburx=435,bbury=171}\caption{{\protect\footnotesize
{(Color online) Schematic setup of the model. The quantum state
initially encoded on the mechanical mode 1 coupled to cavity 1 via
radiation pressure is transferred to the mechanical mode 2 coupled
to cavity 2 via radiation pressure. The state can be transferred
completely without considering leakage rates $\kappa _j$ and
${\gamma _j}$.}}}
\end{figure}

In the interaction picture, the Hamiltonian of the system can be
written as $^{[18-21]}$,
\begin{eqnarray}
H{\rm{ = }}&{\omega _{\rm{m_1}}}b_1^\dag {b_1} + {\omega _{\rm{m_2}}}b_2^\dag {b_2} - {\Delta _1}a_1^\dag {a_1} -{\Delta _2}a_2^\dag {a_2} + {\omega _c}{c^\dag }c \\
&+ g\left( {a_1^\dag c + a_2^\dag c + {a_1}{c^\dag } + {a_2}{c^\dag }} \right) \\
&+ {G_1}\left( {{a_1}b_1^\dag  + a_1^\dag {b_1}} \right) +
{G_2}\left( {{a_2}b_2^\dag  + a_2^\dag {b_2}} \right)\
\end{eqnarray}

where ${\omega _{\rm{m_j}}}$ is the frequency of the mechanical
mode j, $b_j$ is the annihilation operator for the mechanical mode
j, $a_j$ is the annihilation operator for the cavity mode j, $
{\Delta _j}$ is the detuning of the optical drive applied to the
cavity mode j, g is the coupling strength between the cavity and
optical fiber, and $G_j$ is the coupling strength between the
mechanical mode j and cavity j. Through the unitary
transformation, we get the effective Hamiltonian of the system,

\begin{eqnarray}
H'{\rm{ = }}&{\omega _{\rm{m}}}b_1^\dag {b_1} + {\omega _{\rm{m}}}b_2^\dag {b_2} - \left( {{\Delta _1} -\frac{{2{g^2}}}{\Delta }} \right)a_1^\dag {a_1}\\
&- \left( {{\Delta _2} - \frac{{2{g^2}}}{\Delta }} \right)a_2^\dag {a_2}+ \left( {{\omega _c} - \frac{{4{g^2}}}{\Delta}} \right){c^\dag }c \\
&+ \frac{{2{g^2}}}{\Delta }\left( {a_1^\dag {a_2} + {a_1}a_2^\dag } \right)+ {G_1}\left( {{a_1}b_1^\dag  + a_1^\dag{b_1}} \right) \\
&+ {G_2}\left( {{a_2}b_2^\dag  + a_2^\dag {b_2}} \right)\
\end{eqnarray}

here we choose $U = {e^{\frac{g}{\Delta }\left( {a_1^\dag c +
a_2^\dag c - {a_1}{c^\dag } - {a_2}{c^\dag }} \right)}}$, and $H'
= UH{U^\dag }$. For easier calculation, we can choose ${\Delta _1}
= {\Delta _2} =  - \Delta  - {\omega _c}$, at the same time it's
under the case of large detuning conditions, $\Delta  >  > g$. In
the Heisenberg picture, using input-output theory, we obtain the
quantum Langevin equations for the cavity and mechanical mode
operators
\begin{eqnarray}
{{\dot a}_1} =& i\left( {{\Delta _1} - \frac{{2{g^2}}}{\Delta }} \right){a_1} - i{G_1}{b_1} - i\frac{{2{g^2}}}{\Delta}{a_2} - \frac{{{\kappa _1}}}{2}{a_1}\\
& + \sqrt {{\kappa _1}} {a_{in,1}}
\end{eqnarray}

\begin{eqnarray}
{{\dot a}_2} =& i\left( {{\Delta _2} - \frac{{2{g^2}}}{\Delta }} \right){a_2} - i{G_2}{b_2} - i\frac{{2{g^2}}}{\Delta}{a_1} - \frac{{{\kappa _2}}}{2}{a_2}\\
& + \sqrt {{\kappa _2}} {a_{in,2}}
\end{eqnarray}

\begin{eqnarray}
{{{\dot b}_1} =  - i{\omega _{\rm{m}}}{b_1} - i{G_1}{a_1} -
\frac{{{\gamma _1}}}{2}{b_1} + \sqrt {{\gamma _1}} {f_1}}
\end{eqnarray}

\begin{eqnarray}
{{{\dot b}_2} =  - i{\omega _{\rm{m}}}{b_2} - i{G_2}{a_2} -
\frac{{{\gamma _2}}}{2}{b_2} + \sqrt {{\gamma _2}} {f_2}}
\end{eqnarray}

where ${\kappa _j}$ is the damping rate of the cavity mode j,
${\gamma _j}$ is the damping rate of the mechanical oscillator
mode j, ${f_j}$ is the noise operator for the mechanical
oscillator bath j, while ${a_{in,j}}$ describes the quantum noise
of the vacuum field incident on the cavity mode j. Under ideal
conditions, we can choose ${\kappa _1} = {\kappa _2} = {\gamma _1}
= {\gamma _2} = 0$. And only considering the interaction terms, we
get the Langevin equations without dissipative terms.
\begin{eqnarray}
&{{{\dot a}_1} =  - i{G_1}{b_1} - i\frac{{2{g^2}}}{\Delta }{a_2}}\\
&{{{\dot a}_2} =  - i{G_2}{b_2} - i\frac{{2{g^2}}}{\Delta }{a_1}}\\
&{{{\dot b}_1} =  - i{G_1}{a_1}}\\
&{{{\dot b}_2} =  - i{G_2}{a_2}}
\end{eqnarray}

\begin{figure}
\epsfig{file=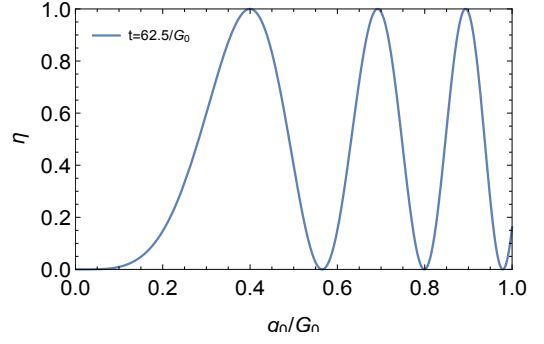, width=7cm,
height=4.5cm,bbllx=0,bblly=0,bburx=263,bbury=175}\caption{{\protect\footnotesize
{(Color online) The state transfer efficiency ${\eta}$ as a
function of ${G_0}/{g_0}$ and for procedure time $t =
{20}/{g_0}$.}}}
\end{figure}

\begin{figure}
\epsfig{file=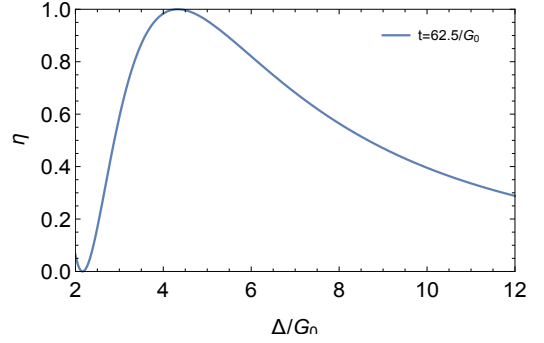, width=7cm,
height=4.5cm,bbllx=0,bblly=0,bburx=263,bbury=175}\caption{{\protect\footnotesize
{(Color online) The state transfer efficiency ${\eta}$ as a
function of ${\Delta}/{g_0}$ and also for the time $t =
{20}/{g_0}$. }}}
\end{figure}

\begin{figure}
\epsfig{file=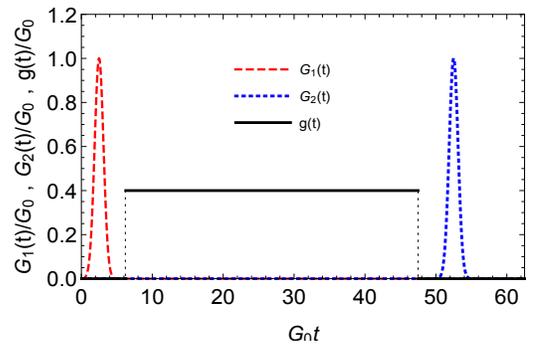, width=7cm,
height=4.5cm,bbllx=0,bblly=0,bburx=263,bbury=175}\caption{{\protect\footnotesize
{(Color online) The time profiles of the optomechanical coupling
${G_1}$ (red dashed curve), ${G_2}$ (blue dashed curve), and the
coupling between cavity and optical fiber $g$ (black solid
curve).}}}
\end{figure}

To analyze the quantum state transfer from mechanical oscillator
mode 1 to mechanical oscillator mode 2, we use the corresponding
classical equations for Eqs. (7)-(10):
\begin{eqnarray}
&{{\alpha _1} =  - i{G_1}{\beta _1} - i\frac{{2{g^2}}}{\Delta }{\alpha _2}}\\
&{{\alpha _2} =  - i{G_2}{\beta _2} - i\frac{{2{g^2}}}{\Delta }{\alpha _1}}\\
&{{\beta _1} =  - i{G_1}{\alpha _1}}\\
&{{\beta _2} =  - i{G_2}{\alpha _2}}
\end{eqnarray}

We assume that only the mechanical oscillator mode 1 has
excitation phonon at the initial moment. To solve the above
equations numerically, we assume ${\beta _1}\left( 0 \right) =
1,{\alpha _1}\left( 0 \right) = 0,{\alpha _2}\left( 0 \right) =
0$,  ${\beta _2}\left( 0 \right) = 0$, and the transfer efficiency
of quantum state as $\eta  = {{{{\left| {{\beta _2}\left( t
\right)} \right|}^2}} \mathord{\left/
 {\vphantom {{{{\left| {{\beta _2}\left( t \right)} \right|}^2}} {{{\left| {{\beta _1}\left( 0 \right)} \right|}^2}}}}
 \right.
 \kern-\nulldelimiterspace} {{{\left| {{\beta _1}\left( 0 \right)} \right|}^2}}}$.

\begin{figure}
\epsfig{file=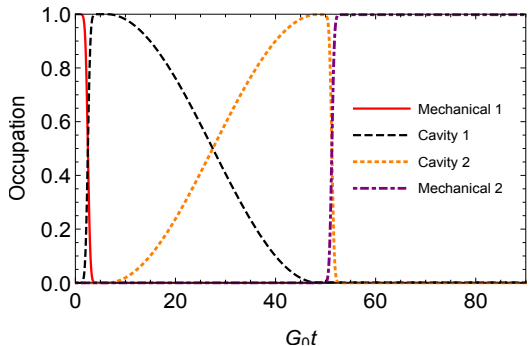, width=7cm,
height=4.5cm,bbllx=0,bblly=0,bburx=263,bbury=175}\caption{{\protect\footnotesize
{(Color online) Occupation number for the four modes. The quantum
state is encoded on the first mechanical mode (red solid curve)
and then transfer to the cavity 1 (black solid curve) and cavity 2
(blue dashed curve). At last, the quantum state is transferred to
the mechanical mode 2 (purple solid curve) In the process, we make
the center of Gaussian pulse at ${t_1}={1}/{g_0}$, ${t_2} =
{10}/{g_0}$ and $s=0.25/{g_0}$.}}}
\end{figure}

Before simulating evolution, we choose a Gaussian pulsed
optomechanical coupling strength of the form ${G_1}\left( t
\right) ={G_0}{e^{{{ - {{\left( {t - {t_1}} \right)}^2}}
\mathord{\left/{\vphantom {{ - {{\left( {t - {t_1}} \right)}^2}}
{2{s^2}}}} \right.\kern-\nulldelimiterspace} {2{s^2}}}}}$,
${G_2}\left( t \right) = {G_0}{e^{{{ - {{\left( {t - {t_2}}
\right)}^2}} \mathord{\left/{\vphantom {{ - {{\left( {t -
{t_2}}\right)}^2}}{2{s^2}}}}\right.\kern-\nulldelimiterspace}
{2{s^2}}}}}$, where ${G_0}$ is the maximum optomechanical coupling
strength, s is the width of the Gaussian pulses, and ${t_i}$ are
the time at which the optomechanical coupling strength is maximum
and the quantum state transfer from one mode to the other occurs.
In addition, the form of the coupling strength between cavity and
optical fiber is $g = {g_0}$ when $t < 9{g_0}$, and if ${t \ge
9{g_0}}$, then $g = 0$.

From Fig.2-4, we obtain the procedure values of the maximum
optomechanical coupling strength ${G_0}$ and detuning ${\Delta}$
which maximize the transfer efficiency ${\eta}$. The values are
${G_0} = 2.5{g_0}$, $\Delta  = 10.5{g_0}$. In such a case, we can
get a high transfer efficiency, just as shown in Fig. 5. When
considering the dissipation ${\kappa}$ and ${\gamma}$, we get FIG.
6. Here the value of dissipations are ${\kappa _1} = {\kappa _2} =
{\gamma _1} = {\gamma _2} = 0.01{g_0}$.

\begin{figure}
\epsfig{file=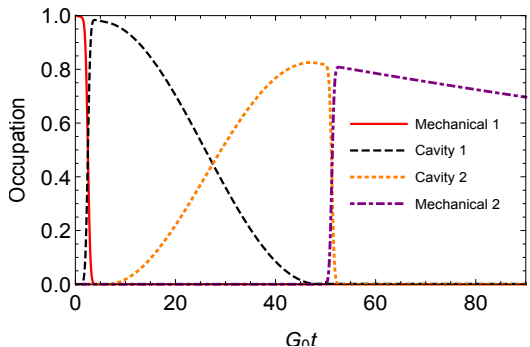, width=7cm,
height=4.5cm,bbllx=0,bblly=0,bburx=263,bbury=175}\caption{{\protect\footnotesize
{(Color online) Occupation number for the four modes. The quantum
state is encoded on the first mechanical mode (red solid curve)
and then transfer to the cavity 1 (black solid curve) and cavity 2
(blue dashed curve). At last, the quantum state is transferred to
the mechanical mode 2 (purple solid curve) In the process, we also
make the center of Gaussian pulse at ${t_1}={1}/{g_0}$, ${t_2} =
{10}/{g_0}$ and $s=0.25/{g_0}$.}}}
\end{figure}

Through analyzing the transfer of a quantum state between two
distant mechanical oscillators which are separately coupled to two
optical cavities, we get the optimal parameter solution for
maximizing the transfer efficiency. In the process of analysis, we
employed time-varying optomechanical coupling strength as a
Gaussian pulsed. Choosing appropriate parameters, we got a high
transfer efficiency, which is very closing to unity.

This work is supported by NSF of China under Grant No. 11305021
and the Fundamental Research Funds for the Central Universities
under Grant Nos 3132015149 and 3032017072.

\end{CJK*}
\end{document}